\documentclass[12pt,a4]{article}
\usepackage[symbol]{footmisc}
\def\mbf#1{\mbox{\boldmath$#1$\unboldmath}}

\font\eufmtex=eufm10

\def\mbf#1{\mbox{\boldmath$#1$\unboldmath}}
\def\Gauss{\mbox{{\eufmtex G}}}
\def\Dij{\mbox{{\eufmtex D}}}
\def\Iij{\mbox{{\eufmtex I}}}
\def\VGot{\mbox{{\eufmtex V}}}

\def\LGot{\mbox{{\eufmtex L}}}

\begin{document}
\begin{center}
{\Large \bf Super Gaussian Self-Consistent method for systems with
a two--body Hamiltonian} \\
\vspace{4 mm}

      Edward G. Timoshenko\footnote{
English translation (2025): E-mail: Edward.Timoshenko@ucd.ie;
Web: http://www.edtim.live
}\footnote{Presently retired}\footnote{
Published in: The Proceedings of "The XVIIth International QFTHEP Workshop," 4-11 September 2003, Samara-Saratov, Russia, Skobeltsyn Institute of Nuclear Physics, MSU, MSU Publishing Co., (M. Dubinin, V. Savrin eds.), pp. 443-449 (2004). 
}

\vspace{4 mm}
Theory and Computation Group,
Centre for Synthesis and Chemical Biology,\\
Conway Institute of Biomolecular and Biomedical Research,
Department of Chemistry, University College Dublin,
Belfield, Dublin 4, Ireland\\
\end{center}

\begin{abstract}
A new kinetic self--consistent method is presented based
on the proposed Gaussian Superposition Principle
for computation of ensemble averaged  observables of a
macromolecule interacting via two--body forces.
The latter leads
to the derivation of a natural functional closure relation for the 3-point
distribution functions (DF), thereby truncating a hierarchy of
kinetic equations obtained from the original Langevin equation.
The resulting Super Gaussian Self--Consistent (SGSC) equations
for the 2-point distribution functions acquire a sufficiently
tractable integro--differential form.
The SGSC theory strives to yield realistic shapes
of various distribution functions for any macromolecule with
a generic Hamiltonian involving 2-body interaction potentials,
both at equilibrium and during kinetics.
\end{abstract}

\section{Introduction}



Models of macromolecules 
are described by fairly complex Hamiltonians
involving both bonded and nonbonded interactions
\cite{CloizeauxBook}. While computer simulations 
\cite{PolSimul}
are proving increasingly powerful in tackling these models, the challenge
of resolving the overall problem of determining the macromolecular 
conformations still remains.
The difficulties with overcoming barriers 
(quasi--nonergodicity), uniformity of sampling the phase space 
for extended polymer chains (entropic suppression), and the shear task of 
reaching satisfactory
equilibration and acceptable sampling statistics all pose severe
problems for equilibrium simulations of macromolecules. 

Moreover, kinetic simulations are further exacerbated 
by the need to limit oneself to ensemble averaging as temporal 
sampling is no longer allowed, the difficulties with adequate 
collective `cluster' motions, and a horrendous
timescale problem brought about by the tiny elementary timestep permitted
(e.g. about 1 fs for Molecular Dynamics), to mention but a few.
There are also various technical limitations and numerical 
inaccuracies intrinsic to each of the standard techniques such as 
Monte Carlo, Stochastic and Molecular Dynamics.

Thus, it is quite evident that an adequate theoretical base is really
needed \cite{SGSC} to interpret and support any simulation data.
Many ingenious theoretical treatments have been developed for polymers at 
equilibrium over the years\cite{CloizeauxBook,PolTheor}. 
However, in the main, in order to be amenable to analytical analysis,
they often have to be either based on rather simplistic models of polymers, 
or to involve certain approximations with profound consequences for the results.
Much less developed and few are 
the nonequilibrium methods, which perform satisfactorily for describing 
conformational changes of macromolecules. It is often the case, therefore, 
that a great deal of the description in the polymer area is only qualitatively 
accurate, but lacking the firm quantitative rigour of other theoretical physical 
disciplines. 

Fundamentally, for the description of kinetics 
of conformational changes
of a macromolecule, one would like to be able to reliably deduce its most
detailed statistical--mechanical properties as a function of time given
the system's Hamiltonian in terms of a classical force field.
As regards the equilibrium state of the macromolecule, it should 
be simply recoverable as the stationary limit in time of the correct kinetic
equations.

An approach to the problem based on the Gaussian self--consistent (GSC)
method has been independently developed and applied in somewhat different
formulations by our Group \cite{GSC} and
others (see e.g. Refs. \cite{GanazzoliGSC,OrlandPitard}
and further references therein).
Historically, this technique had a predecessor in the Gaussian
variational approach for a linear homopolymer
chain at equilibrium independently proposed by J.~des Cloizeaux 
\cite{Cloizeaux-var} and
S.~Edwards {\it et al} \cite{Edwards-var}. Further progress has
been made in the years which followed \cite{Allegra} with
generalisations to polymers with different chain architectures
\cite{Raos-94}, such as e.g. dendrimers \cite{DendriHomo}, 
chain stiffness \cite{Torus},
copolymers \cite{Ganazzoli-00}, and multiple chains \cite{Analysis}.
However, the method becomes quantitatively inaccurate for nonideal systems
even for these observables and, clearly, it is unable to yield 
realistic radial dependencies of the densities and distribution functions.
In particular, the GSC results suffer from an overestimated expansion 
of the coil in the good solvent, leading to a too large swelling exponent 
value as the degree of polymerisation increases,
whereas an underestimated energy of the globule in the poor solvent.
The cause of both deficiencies is quite apparent when comparing the 2-point 
DF of the GSC and MC.
The 2-point DF of GSC has a smooth Gaussian shape extrapolating the true
nonmonotonous DF of the nonideal systems. The latter functions 
have a pronounced power law decrease at the origin, 
the so--called `correlation hole',
for the coil, whereas an oscillating liquid--like shape with a pronounced
peak at the `first solvation shell' radius for the globule \cite{CorFunc}.

\section{The new SGSC method}

The intra--chain correlation function is defined as,
\begin{equation}
g^{(2)}_{ij}(r) \equiv \left\langle \delta(\mbf{x}_i - \mbf{x}_{j}-\mbf{r}) 
\right\rangle = \frac{1}{4\pi r^2} \left\langle \delta(|\mbf{x}_{i} - \mbf{x}_{j}|-r) \right\rangle,
\end{equation}
We also would need the mean--squared distances between monomers,
\begin{equation}
\Dij_{ij}\equiv\frac{1}{3}\langle (\mbf{x}_i-\mbf{x}_j)^2 \rangle=
\frac{4\pi}{3}\int_0^{\infty} r^3 dr\,{\cal G}_{ij}(r),
\quad {\cal G}_{ij}(r)\equiv r\, g^{(2)}_{ij}(r).
\end{equation}
Next we introduce
notations for pairs of indices for brevity,
\begin{equation}
ij \to 1, \quad jk \to 2, \quad ik \to 3 , \quad ji \to \breve{1}.
\end{equation}

We proceed from the coarse--grained Langevin
equation for the atom coordinates $\mbf{x}_i(t)$,
\begin{equation}
\zeta\frac{d\mbf{x}_i}{dt}=-\frac{\partial H}
{\partial \mbf{x}_i}+\mbf{\eta}_i.
\end{equation}
We can restrict ourselves by a 2-body Hamiltonian $H$ 
and a Gaussian (white) noise,
\begin{equation}
H=\sum_{j<k}V_{jk}(|\mbf{x}_j-\mbf{x}_k|), \qquad
\left\langle \eta_i^{\alpha}(t)\,\eta_j^{\beta}(t') \right\rangle=
2 k_B T \zeta \, \delta^{\alpha\beta}\,\delta_{ij}\,\delta(t-t').
\end{equation}

The exact equation can be derived with help of Fluctuation--Dissipation Theorem,
\begin{eqnarray}
\zeta\frac{d g_{ij}^{(2)}(\mbf{r})}{dt}&=&
2k_B T \frac{\partial^2 g_{ij}^{(2)}(\mbf{r})}{\partial \mbf{r}^2} 
+ \left\langle
\frac{\partial G^{(2)}_{ij}(\mbf{r})}
{\partial \mbf{r}}
\left(\frac{\partial H}{\partial \mbf{x}_i}
-\frac{\partial H}{\partial \mbf{x}_j}\right)\right\rangle. 
\end{eqnarray}
This eq. involves 3-point DF and leads to a kinetic hierarchy
analogous to that of BBGKY
(Bogoliubov, Born, Green, Kirkwood and Yvon)
derived from the Liouville eq. \cite{Bogol,HansenMcDonald},
\begin{eqnarray}
\label{HEq}
&&I_{ij}\equiv \left\langle
\frac{\partial G^{(2)}_{ij}(\mbf{r})}
{\partial \mbf{r}}
\left(\frac{\partial H}{\partial \mbf{x}_i}
-\frac{\partial H}
{\partial \mbf{x}_j}\right)\right\rangle, \quad
I^{(dir)}_{ij}=\frac{2}{r^2}\frac{\partial}{\partial r}\biggl( 
r^2 \frac{\partial V_{ij}(r)}{\partial r}\,g^{(2)}_{ij}(r)
\biggr),
\ \\
&& g_{ijj}^{(3)}(\mbf{r},\mbf{r}')=g^{(2)}_{ij}(r)\,\delta(\mbf{r}'), 
\qquad 
g_{iji}^{(3)}(\mbf{r},\mbf{r}')=g^{(2)}_{ij}(r)\,\delta(\mbf{r}+\mbf{r}').
\\
&&I_{ij}^{(indir)}= \sum_{k\not=i,j} \int d\mbf{r}'\,
\left ( V_{jk}(\mbf{r}')
\frac{\partial^2 g^{(3)}_{ijk}(\mbf{r},\mbf{r}')}
{\partial\mbf{r}\partial\mbf{r}'}
- V_{ik}(\mbf{r}')
\frac{\partial^2 g^{(3)}_{jik}(-\mbf{r},\mbf{r}')}
{\partial\mbf{r}\partial\mbf{r}'} \right). 
\end{eqnarray}

For any observable $A$,
we formulate the Gaussian Superposition Principle (GSP),
\begin{equation}
\!\!
\biggl\langle A(\{\mbf{x}_i(t) -\mbf{x}_j(t)\}) \biggr\rangle =
\prod_{i<j}\int d\mbf{y}_{ij}\, 
f^t_{ij}(|\mbf{y}_{ij})\, 
\left\langle A(\{\mbf{x}_i^{(0)}(t) -\mbf{x}_j^{(0)}(t)
-\mbf{y}_{ij}\sqrt{{\cal D}_{ij}(t)}\}) 
\right\rangle_0, 
\end{equation}
where $\langle \ldots \rangle_0$ denotes the averaging over the Gaussian
noise via the GSC method,
\begin{equation}
\zeta\frac{d\mbf{x}_i^{(0)}}{dt}=-\frac{\partial H^{(0)}}
{\partial \mbf{x}_i^{(0)}}+\mbf{\eta}_i, \qquad
H^{(0)} = \sum_{j<k} K_{jk} \mbf{x}_j^{(0)}\,\mbf{x}_k^{(0)},
\end{equation}
The variables $\mbf{y}_{ij}$ are  dimensionless
and ${\cal D}_{ij}$ are the MS distances
of the GSC method,
\begin{equation}
{\cal D}_{ij}(t)\equiv \frac{1}{3}\biggl\langle \left( 
\mbf{x}_i^{(0)}(t)-\mbf{x}_j^{(0)}(t) \right)^2
\biggr\rangle_0. 
\end{equation}
The functions $f_{ij}$ for any $ij$ satisfy the normalisation conditions,
\begin{equation}
\int d\mbf{y}_{ij}\, f_{ij}^t(|\mbf{y}_{ij}|)=1.
\end{equation}

In application to $g^{(2)}_{ij}$, we can hence produce the decomposition,
\begin{equation}
g^{(2)}_{ij}(r,t) =  \int d\mbf{y}\,
(2\pi {\cal D}_{ij}(t))^{-3/2}
\exp\left( -\frac{\left(\mbf{r}-\mbf{y}\sqrt{{\cal D}_{ij}(t)}\right)^2}
{2{\cal D}_{ij}(t)}\right)
\,f_{ij}^t(y). \label{g2Super}
\end{equation}
Next, let us introduce the rescaled variables and functions,
\begin{equation}
 \hat{r}_{ij}\equiv \frac{r}{\sqrt{{\cal D}_{ij}}}, \quad
g^{(2)}_{ij}(r) \equiv \frac{\hat{g}^{(2)}_{ij}(\hat{r}_{ij})}
{{\cal D}_{ij}^{3/2} }, 
\quad y\,f_{ij}(y)\equiv {\cal F}_{ij}(y), \quad 
\hat{r}_{ij}\hat{g}^{(2)}_{ij}(\hat{r}_{ij})\equiv 
\hat{{\cal G}}_{ij}(\hat{r}_{ij}).
\end{equation}
We can rewrite above Eq. as an equivalent 1-D decomposition,
\begin{equation}
\hat{{\cal G}}_{ij}(\hat{r}_{ij})=\int_{-\infty}^{\infty} dy\,
\frac{\exp\left( -\frac{(\hat{r}_{ij}-y)^2}{2}\right)}{(2\pi)^{1/2}}
\,{\cal F}_{ij}(y).
\end{equation}
Mathematically, this Eq, is the Gauss transformation (GT),
\begin{eqnarray}
G(x) &=&  \Gauss[F](x) \equiv
\int_{-\infty}^{\infty}dy\, 
\frac{\exp\left( -\frac{(x-y)^2}{2}\right)}{(2\pi)^{1/2}}\,F(y)
=\exp\biggl(\frac{1}{2}\frac{d^2}{dx^2} \biggr)\,F(x),
\\
F(x) &=& \Gauss^{-1}[G](x) =  
 \exp\biggl(-\frac{1}{2}\frac{d^2}{dx^2} \biggr)\,G(x). 
\end{eqnarray}

Similarly, the GSP of the 3-point DF gives the expression,
\begin{eqnarray}
&& g^{(3)}_{ijk}(\mbf{r},\mbf{r}') =
\int \int d\mbf{y} d\mbf{y}'\, 
\frac{f_{ij}(y)\,f_{jk}(y')}{\left((2\pi)^2\,\Delta_{ijk}\right)^{3/2}}\,
\exp\biggl(-\frac{1}{2\Delta_{ijk}}\biggl[
(\mbf{r} -\mbf{y}\sqrt{D_{ij}})^2{\cal D}_{jk} \biggr.\biggr.
\nonumber\\ &&
\biggl.\biggl.
-2 (\mbf{r}-\mbf{y}\sqrt{D_{ij}})
(\mbf{r}'-\mbf{y}'\sqrt{D_{jk}}){\cal D}_{ij,jk}
+(\mbf{r}'-\mbf{y}'\sqrt{D_{jk}})^2{\cal D}_{ij}
\biggr]\biggr),
\end{eqnarray}
where $\Delta_{ijk}> 0$ and the 
correlation coefficients ${\cal D}_{ij,jk}$ are,\\
\begin{eqnarray}
\Delta_{ijk} & \equiv  & {\cal D}_{ij}\,{\cal D}_{jk}
-{\cal D}_{ij,jk}^2,  \\
{\cal D}_{ij,jk} & \equiv  & 
\frac{1}{3}\biggl\langle \left( \mbf{x}_i^{(0)} -\mbf{x}_j^{(0)} \right)
\left( \mbf{x}_j^{(0)} -\mbf{x}_k^{(0)} \right) \biggr\rangle_0
 = -\frac{1}{2}
\left({\cal D}_{ij}+{\cal D}_{jk}-{\cal D}_{ik}\right).
\end{eqnarray}
From above Eq. we deduce the 
following closure relation,
\begin{equation}
g^{(3)}_{ijk}(\mbf{r},\mbf{r}')=\exp\biggl(
{\cal D}_{ij,jk}\frac{\partial^2}{\partial \mbf{r}\,\partial \mbf{r}'}
\biggr)\,g_{ij}^{(2)}(r)\,g_{jk}^{(2)}(r').
\end{equation}
Properties of this SGSC closure relation include:
\begin{eqnarray}
&& \Dij_{ij,jk}={\cal D}_{ij,jk}=\frac{1}{3}\int\int d\mbf{r}d\mbf{r}'\, (\mbf{r}\mbf{r}')\,
g^{(3)}_{ijk}(\mbf{r},\mbf{r}').
\\
&&\int d\mbf{r}'\, g^{(3)}_{ijk}(\mbf{r},\mbf{r}')= g_{ij}^{(2)}(r),
\qquad \int d\mbf{r}\, g^{(3)}_{ijk}(\mbf{r},\mbf{r}')= g_{jk}^{(2)}(r').
\end{eqnarray}
Moreover, the proposed SGSC closure satisfies the correspondence
condition, i.e. it recovers the exact GSC closure relation
if 
$g^{(2)\,Gau}_{ij}(r)=(2\pi {\cal D}_{ij})^{-3/2}\exp(-r^2/(2{\cal D}_{ij}))$.
Note that to obtain the latter distribution we simply have to choose 
$f_{ij}^{Gau}(\mbf{y})=\delta(\mbf{y})$.

An important feature of this theory is the
1-D reduction of the 3-point DF,
\begin{equation}
{\cal I}_{12}\equiv\int d\stackrel{\triangle}{\mbf{r}}{\!\!}'\,
\frac{\partial^2 g^{(3)}_{12}(\mbf{r},\mbf{r}')}{\partial \mbf{r}
\,\partial \mbf{r}'}=
\frac{2\pi}{rr'}\, \Xi\biggl({\cal D}_{1,2}
\frac{\partial}{\partial r}\frac{\partial}{\partial r'}\biggr)
\frac{\partial}{\partial r}\frac{\partial}{\partial r'}\,
rg_1^{(2)}(r)\,r'g_2^{(2)}(r'),
\end{equation}
where $\Xi(z) \equiv \int_{-1}^1 da\,a\,e^{a\,z}$.
The resulting SGSC equation for the 2-point DF functions are,
\begin{eqnarray}
&& \zeta \frac{d {\cal G}_{1}(r)}{dt} =  
2k_B T \frac{\partial^2 {\cal G}_{1}(r)}{\partial r^2}  
+\frac{2}{r}\frac{\partial}{\partial r}\biggl( 
r\frac{\partial V_{1}(r)}{\partial r}\,{\cal G}_{1}(r)
\biggr)  \\
&&+ 2\pi \sum_{k\not=i,j} \int_0^{\infty}dr'\, \biggl[ r'\, V_{2}(r')\,
\Xi\biggl({\cal D}_{1,2} \frac{\partial}{\partial r}
\frac{\partial}{\partial r'}\biggr) 
\frac{\partial}{\partial r}\frac{\partial}{\partial r'}
{\cal G}_{1}(r)\,{\cal G}_{2}(r') 
 + \bigl((1,2)\leftrightarrow (\breve{1},3)
\bigr) \biggr].
\nonumber
\end{eqnarray}

It would appear that the composition operator 
$\Xi$ generates a seemingly intractable 
series involving the radial derivatives of all orders. However,
this operator is closely related to the translation operator. 
By expressing the potential 
$r V_2(r)$ as 
the Laplace transform of its original one can rewrite 
this as,
\begin{equation}
\label{UindirTwo}
U^{(indir)}_{1,2}=\int_{-\infty}^{\infty}\tau\,d\tau\,
\hat{{\cal G}}_1^{[1]}(\hat{r}_1+\Gamma\tau)\int_0^{\infty}d\hat{r}_2\,
\VGot_2(\hat{r}_2,|\tau|,{\cal D}_2)\,\hat{{\cal G}}_2^{[1]}(\hat{r}_2),
\end{equation}
where 
the derivative $\hat{{\cal G}}_1^{[1]}$
is understood to be an even function
and we have introduced,
\begin{equation}
\label{FauDef}
\VGot_2(\hat{r}_2,|\tau|,{\cal D}_2)\equiv \int_{|\tau|}^{\infty}\frac{ds}{s^2}
\frac{\LGot^{-1}[rV_2](s/\sqrt{{\cal D}}_2)}{{\cal D}_2}\,
\exp\left( -s\hat{r}_2 \right).
\end{equation}

Since $\Iij_{ij}$ is not an observable as it can not
be represented via an average of some function of monomer
coordinates, it is not clear how to obtain an evolution
equation for it. The same difficulty exists for ${\cal D}_{ij}$,
which we still need to know in order to express the correlation
coefficients ${\cal D}_{ij,jk}$
involved in the closure relation.
The most obvious thought is to require the Gaussian quantities 
${\cal D}_{ij}$ to satisfy a quasi--Gaussian equation similar
to Eq. (17) in Ref. \cite{Conftra}, which involves the
derivatives of the instantaneous mean energy,
\begin{eqnarray}
\frac{\zeta}{2} \frac{d {\cal D}_{ij}}{dt} & = & 2k_B T +
\sum_k ({\cal D}_{ik} -{\cal D}_{jk})(W_{ik}-W_{jk}), \label{Dg1}\\
W_{ik}(t) & \equiv & -\frac{2}{3}\frac{\partial {\cal E}}{\partial {\cal D}_{ik}(t)},
\qquad {\cal E} \equiv \langle H \rangle =4\pi \sum_{j'<k'}
\int_0^{\infty} r^2dr\,V_{j'k'}(r)\,g^{(2)}_{j'k'}(r).  \label{Dg2}
\end{eqnarray}
Here we use the full mean energy as the integral of
the Gaussian function with the 2-body potential would diverge.
After a lengthy algebra  the final
two equations become,
\begin{eqnarray}
\frac{\zeta}{2}\frac{d\Dij_1}{dt}  &=& X_1,   
\qquad \frac{\zeta}{2}\frac{d{\cal D}_1}{dt} = Y_1,
\label{DIeq}\\
X_1 & \equiv &  2k_B T +2 A_1 + \sum_{k\not= i,j}({\cal D}_{1,2} B_2 
+ {\cal D}_{\breve{1},3} B_3), \\
Y_1 & \equiv&  2k_B T + 2 A_1 - \sum_{k\not=i,j}({\cal D}_{1,2} 
\frac{A_2}{{\cal D}_2}
+ {\cal D}_{\breve{1},3} \frac{A_3}{{\cal D}_3}), \label{XY}\\
A_1 & \equiv &  \frac{4\pi}{3} \int_0^{\infty} dr\,
V_1(r)
\frac{\partial (r^2{\cal G}_1)}{\partial r}, \qquad
B_1 \equiv \frac{4\pi}{3}\int_0^{\infty} dr\,r\,
V_1(r)
\frac{\partial^2 {\cal G}_1}{\partial r^2}.
\label{AB}
\end{eqnarray}

\section{Conclusion}
Here I have shown how one can extend the Gaussian 
self--consistent (GSC) approach for describing conformational changes 
of a macromolecule so that the Gaussian limitation is surpassed. 
Obviously, a Gaussian shape of the monomer---monomer 
distribution functions (DFs)
is correct only for an ideal molecule characterised by a set of 
arbitrarily connected harmonic springs. Such Gaussian chain, or more 
generally a Gaussian network, being exactly solvable,
plays an important r\^{o}le similarly to other ideal systems in 
Statistical Mechanics.
The GSC method is based on finding a trial Gaussian network,
involving virtual harmonic springs between all of the monomers,
which fits best the mean--squared (MS) inter--monomer distances
with respect to the true system Hamiltonian. Such a trial network
is described by a quadratic trial Hamiltonian, and hence by a linear
trial stochastic ensemble associated with it, leading to an unrealistic
Gaussian shape of DFs.

Clearly, any physically reasonable shape of DF can be represented as
a linear superposition of shifted Gaussian functions of the same
width, which is known mathematically as Gauss transformation 
(GT).
Therefore, given the true shape of DF, one can invert the GT in order to find
the superposition amplitudes $f_{ij}^t(r)$.
The idea of the SGSC method is to find such 
amplitudes in a self--consistent manner proceeding from the nonlinear Langevin
equation by optimising the 2-point DFs, similar to how the GSC method 
optimises the MS distances between monomers. This is accomplished as follows. 
First, one starts by rigorously deriving a hierarchy of equations for DFs
from the Langevin equation similarly to the BBGKY hierarchy obtained from
the Liouville equation \cite{Bogol,HansenMcDonald}. Second,
we write the 3-point DF as a double linear superposition
of the reference Gaussian 3-point DFs with the same
decomposition amplitudes as for the 2-point DF, which, in fact,
constitutes a natural functional closure relation for the
3-point function in terms of the 2-point functions. Such a closure
truncates the infinite hierarchy and produces
a closed set of the SGSC equations for the 2-point DFs. 
Remarkably, this generic closure relation,
which is valid for any 2-body potential,
happens to possess all of the properties required by Statistical Mechanics.
Furthermore, it turns out to be also mathematically well tractable, leading
in a {\it tour de magie} to the 1-dimensional reduction
and, consequently, to an elegant final equation by
exploiting the intimate relation of the closure to the translation operator.
We know that the latter operator is of great significance in
Quantum Mechanics. In particular, the symbol formulation of Quantum 
Mechanics, 
which is expressed in terms of functions over the classical 
phase space, has the convolution operator
for two observables, which looks very similar to the SGSC closure
relation.

If we extend the above superposition method to any observable in our
theory, we would naturally arrive to the general form of the Gaussian
Superposition Principle (GSP), which expresses any observable
of the SGSC theory as a linear superposition of the observables
obtained from the reference Gaussian (GSC) method. One can ask perhaps
how the GSP corresponds to the Superposition Principle in Quantum 
Mechanics, which is considered to be one of its cornerstones.
Any Quantum Mechanical system is described 
by a linear equation in the Hilbert space and
thus superpositions of elementary solutions emerge
as a simple consequence of linearity. 
The current application of the superposition idea to a highly
nonlinear Langevin equation is much more subtle. In fact, what we 
do here is attempting
to construct a good approximation for ensemble averaged observables
by expressing them via a linear superposition of the elementary approximants
obtained from a reference Gaussian theory. This is done
by determining the decomposition amplitudes, 
which optimise the 2-point DFs with respect
to the true nonlinear Langevin equation. Yet, despite this rather
different context, the formal analogies with Quantum Mechanics
in various parts of the present formalism are quite striking
and unforeseen.

Interestingly, the final SGSC equation is a parabolic type
integro--differential equation, which has terms 
analogous to those of the Fokker--Planck equation
\cite{Risken}: the entropic
term is similar to the diffusive part, whereas the direct interactions term is
similar to the drift part of the Fokker--Planck equation.
The remaining indirect interactions term involves an integral
bearing some similarity to the collision integral 
in the Boltzmann\cite{BoltzmannEq} and, particularly, in the Landau
equations.
We know that the latter integral equations
have been very successful in describing a rich physical behaviour
of various complex systems, 
but these equations are generally difficult to solve analytically. 
One would hope that the proposed SGSC equation could play 
a fundamental r\^{o}le for the kinetics of conformational
changes of a macromolecule in solution over long timescales,
akin to the r\^{o}le played by the Boltzmann equation for 
the kinetics of gas phase.

\section*{Acknowledgments}

I am grateful to a number of colleagues and collaborators
for most interesting discussions on this matter over the years, 
but as the list would go on quite long, I would only especially
thank Dr Yuri A. Kuznetsov for all his support and help.


%

\end{document}